\documentclass[10pt,a4paper,journal,english]{IEEEtran}

\usepackage[T1]{fontenc}
\usepackage[utf8]{inputenc}

\usepackage{babel}
\usepackage[subrefformat=parens,skip=6pt]{subcaption}
\usepackage{graphicx}
\usepackage{rotating}
\usepackage{multirow}
\usepackage{amscd}
\usepackage[centertags,reqno]{amsmath}
\usepackage{amssymb}
\usepackage{theorem}
\usepackage{paralist}
\usepackage{verbatim}
\usepackage{balance}
\usepackage{cite}
\usepackage[multiple]{footmisc}
\usepackage{listings}
\usepackage{xcolor}
\usepackage{url}
\usepackage{newverbs}

\newverbcommand{\cverb}{}{}
\newcommand{\ctext}[1]{\texttt{#1}}

\newenvironment{singlespace}
   {}
   {}

\makeatletter
\AtBeginDocument{%
  \let\c@figure\c@lstlisting
  
  \let\ftype@lstlisting\ftype@figure 
}
\makeatother

\lstnewenvironment{lstcommand}[2]
   {\lstset{float=t,language=bash,caption=#2,label=#1}}
   {}

\lstnewenvironment{lstcode}[2]
   {
   \lstset{float=t,language=C++,caption=#2,label=#1}}
   {}

\allowdisplaybreaks

\newcommand{\field}[1]{\ensuremath{\mathbb{#1}}}
\newcommand{\gf}[1]{\ensuremath{\field{F}_#1}}

\newcommand{\dB}{\ensuremath{\,\mathrm{dB}}}

\newlength\figwidth
\newcommand{\insertfig}[3]{%
   \begin{figure}[!t]
      \centering
      \includegraphics[width=\figwidth]{#2}
      \caption{#3}
      \label{#1}
   \end{figure}}

\newcommand{\inserttabd}[4]{%
   \begin{table*}[!t]
      \begin{minipage}{\textwidth}
         \caption{#3}
         \label{#1}
         \centering
         \renewcommand{\arraystretch}{1.15}
         \begin{tabular}{#2}
            \hline
            #4
            \hline
         \end{tabular}
      \end{minipage}
   \end{table*}}

\newcommand{\insertsys}[3]{%
   \begin{singlespace}
   \lstset{language=,morecomment=[l]\#}
   \lstinputlisting[float,label=#1,
      caption=#3]
      {#2}
   \end{singlespace}}

\title{SimCommSys: Taking the errors out of\\ error-correcting code simulations}

\author{%
   Johann~A.~Briffa
   and~Stephan~Wesemeyer%
   \thanks{%
      Manuscript submitted February 20, 2014;
      accepted May 12, 2014 for publication in
      the IET Journal of Engineering.
      }%
   \thanks{%
      J.~A.~Briffa and S.~Wesemeyer are with the
      Dept.\ of Computing, University of Surrey, Guildford GU2 7XH, England.
      Email: j.briffa@surrey.ac.uk
      }%
   }

\begin{document}

\maketitle

\begin{abstract}
In this paper we present SimCommSys, a Simulator of Communication Systems
that we are releasing under an open source license.
The core of the project is a set of C++ libraries defining communication
system components and a distributed Monte Carlo simulator.
Of principal interest is the error-control coding component, where various
kinds of binary and non-binary codes are implemented, including turbo, LDPC,
repeat-accumulate, and Reed-Solomon.
The project also contains a number of ready-to-build binaries implementing
various stages of the communication system (such as the encoder and decoder),
a complete simulator, and a system benchmark.
Finally, SimCommSys also provides a number of shell and python scripts to
encapsulate routine use cases.
As long as the required components are already available in SimCommSys,
the user may simulate complete communication systems of their own design
without any additional programming.
The strict separation of development (needed only to implement new components)
and use (to simulate specific constructions) encourages reproducibility of
experimental work and reduces the likelihood of error.
Following an overview of the framework, we provide some examples of how
to use the framework, including the implementation of a simple codec, the
specification of communication systems and their simulation.
\end{abstract}



\section{Introduction}
\label{sec:introduction}

Error-correcting codes are everywhere, from CDs (using Reed-Solomon codes),
the now interstellar space communication between NASA and the Voyager 1 probe
(using a convolutional code concatenated with a Golay code), to current
terrestrial HDTV broadcasts (DVB-T2, using concatenated LDPC and BCH codes).
Shannon's seminal paper \cite{shan48} showed that almost error-free
communication is possible provided the rate of the information transmitted
is below the capacity of the channel.
Unfortunately, the proof is based on a probabilistic argument and does not
provide a way of constructing codes that achieve the channel capacity while
still being practical to encode and decode.
Since then, the error-correction code community has developed a plethora of
different codes and associated encoders and decoders.
Modern codes, such as turbo \cite{berr93}, LDPC \cite{mack97} and polar
codes \cite{arikan2009jit}, have come very close to the channel capacity
while still having encoders and decoders `simple' enough to be useful.

As new codes continue to be developed whose performance need to be evaluated,
most researchers in this field will at some point have written a simple test
and performance measurement harness.
To do this, it is likely they will have needed to implement or find at least:
\begin{inparaenum}[\em a)]
\item a fast finite field representation for their code alphabet,
\item vector and matrix representations capable of handling finite fields
for their encoder and decoder,
\item a channel implementation to simulate the transmissions of codewords,
\item a framework that glues together the encode, transmit, decode steps for
a large number of codewords and computes the symbol error rate (SER) and frame
error rate (FER), and
\item for more complex systems, a way of using this framework on a cluster
to speed up the processing time.
\end{inparaenum}
While being able to code these things is arguably a good learning experience
for any new researcher, most of these implementation tasks are peripheral
to the main research problem of designing codes.

These elements are needed to test the code but should be independent of the
code and remain unchanged across codes.
In fact, researchers having to implement these components separately is a likely
source of additional errors which might hide problems in the performance of the
designed code.
For example, a channel which introduces less noise than required for a specific
signal-to-noise ratio (SNR) might boost the claimed performance of a code.
Consequently, having a framework of well-tested components that provide all
the required features allows the researcher to concentrate on code design
and testing it.
It also provides other researchers with a means to reproduce the results
more easily.

Current choices for researchers are limited, and each option has its caveats.
Perhaps the most popular solution, MATLAB \cite{matlab} and its Communications
Toolbox includes implementations of a very wide range of current schemes.
However, it comes at a considerable financial cost, particularly for parallel
computation, limiting its availability to researchers.
Furthermore, its usual workflow requires the user to write scripts to describe
a given system, making it easy to introduce logical errors.
The interpreted nature of the language also makes the implementation of novel
code constructions inefficient, unless one makes use of the MATLAB to C/C++
interface (which requires considerable technical skill).
An alternative open-source solution exists in the form of IT++
\cite{lsrsc2006its, cristea2010simutools}, a library of mathematical, signal
processing and communication system components.
While this library includes implementations of most current schemes, it
is designed to be used by programmers, and the implementation of a system
requires the user to write C++ code.
Notably, the library does not include a simulation framework, so that the user
is responsible to collect results and decide when a simulation has converged.

Our Simulator of Communication Systems (SimCommSys) C++ framework helps
address all these issues.
It runs on both Windows and Linux and can also use NVIDIA GPUs using CUDA
\cite{cuda-pg-50}.
It has been developed over more than 15 years by the first author and was
used successfully in a number of papers \cite{briffa13jcomml, bb11isit,
bs08isita, bs08istc, bb02jcom}.
The source code has been released under the GNU General Public Licence (GPL)
version 3 (or later) and can be found, together with its documentation, at:\\
\url{https://github.com/jbresearch/simcommsys}.

Researchers are provided with a host of existing codecs, channels, modulators
and different performance measures to gather results quickly.
Furthermore, the implementation is highly modular and every component is
designed around simple-to-use interfaces which make it straightforward
to extend the framework with new codecs as well as other components like
channels, modulators, etc.\ as required.
Particular attention is also paid to the correctness of implementation
and verifiability, for increased confidence in the results obtained.
Internal checks are performed at multiple levels, and are available to the
user through the debug build.

The rest of this paper is organized as follows.
In Section~\ref{sec:framework} we describe the structure of the framework
in more detail, highlighting the main components of interest.
This is then followed by some examples of how to use SimCommSys
in Section~\ref{sec:using} and how to extend the framework in
Section~\ref{sec:extending}.
We conclude with an invitation to other researchers to use and contribute
to the project.


\section{SimCommSys: The framework}
\label{sec:framework}

\subsection{The communication system model}

Fig.~\ref{fig:comm_model} shows a simplified version of the communication
system model that is supported by our framework.
\insertfig{fig:comm_model}{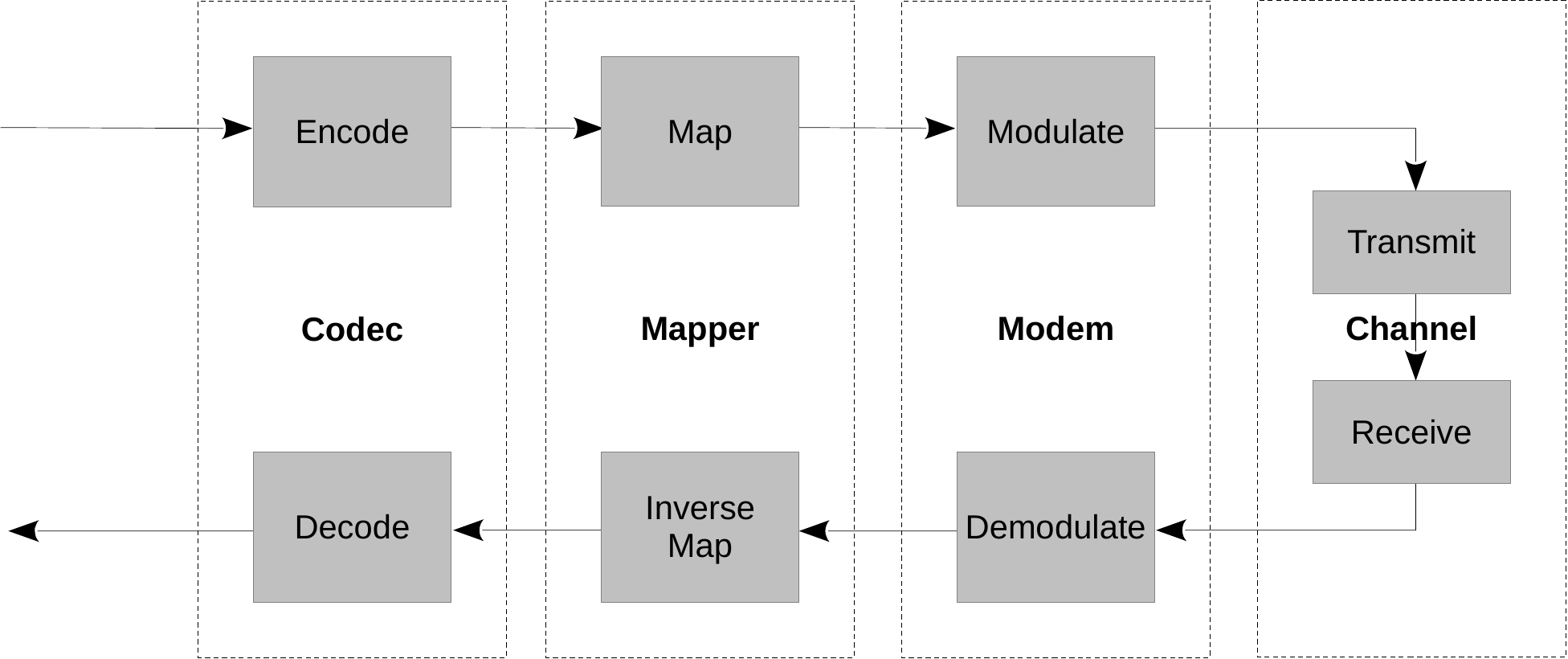}
   {The SimCommSys communication system model}
The codec block represents the code that is to be tested, and consists of
an encoder and its corresponding decoder.
In order to create a new code, the user simply needs to inherit from
the abstract class \cverb|codec| or, if the code can provide soft output,
\cverb|codec_softout|.
In turn this inherits from the \cverb|codec| class; it is provided for
convenience, and also allows iterative decoding between the codec and modems
that support this.

The remaining components are straightforward to explain.
After encoding the information with the encoder, it is the responsibility
of the mapper component to translate, if required, the output symbols of the
encoder to the symbols that can be modulated by the chosen modulation scheme.
For example, this translation is necessary when the code alphabet is over
$\gf{q}$ with $q=2^k$ for some $k>1$, and the channel modulation scheme
is binary.
The mapper can also be used to interleave the codec output and to puncture
it to increase the code rate.
Obviously, at the receiving end any interleaving or puncturing needs to be
undone and the received symbols need to be mapped back to the alphabet that
the codec understands.

The modem is responsible for translating the abstract symbols to their
equivalent channel representation.
A number of different modems are available, including commonly used
ones for channels with a signal-space representation.
These include $M$-ary phase-shift keying (PSK) with variable $M$ (including
$M=2$ for BPSK, etc.) and quadrature amplitude modulation (QAM).
Null modems are also available for abstract channels.

Finally, the channel represents the medium over which the modulated signal
is transmitted and exposed to noise or corruption.
Again a number of channels are available, including the commonly used
additive white Gaussian noise (AWGN) channel.
Asbtract channels are also available, including the $q$-ary erasure channel,
$q$-ary symmetric channel, and also channels with insertion, deletion and
substitution errors.
In the case of such synchronization error channels, the length of the
transmitted sequence (i.e.\ entering the channel) might not be the same as
that of the received sequence.

A summary of the principal communication system components available in the
SimCommSys code base is given in Table~\ref{tab:commsys}.
\inserttabd{tab:commsys}{ccp{4.75in}}
   {Summary of the principal communication system components available in
   the code base.}{
   \emph{Base} & \emph{Class} & \emph{Description} \\
   \hline
   \ctext{codec}
   & \ctext{reedsolomon} & Reed-Solomon code over $\gf{q}$ of length $n \in \{q,q-1\}$ and dimension $1<k<n-1$ with Berlekamp decoder \\
   & \ctext{ldpc} & LDPC code over $\gf{q}$ of length $n$ and dimension $m$ \\
   & \ctext{mapcc} & Convolutional code with BCJR decoder \cite{bahl74} \\
   & \ctext{repacc} & Repeat-Accumulate code with BCJR decoder \\
   & \ctext{sysrepacc} & Systematic Repeat-Accumulate code with BCJR decoder \\
   & \ctext{turbo} & Parallel concatenated convolutional code with variable interleavers and BCJR decoder \\
   & \ctext{memoryless} & Simple mapping, with or without repetition \\
   & \ctext{uncoded} & Uncoded transmission (output is copy of input) \\
   & \ctext{codec\_multiblock} & Meta-codec that concatenates a number of blocks of the underlying codec (for interleaving across blocks) \\
   & \ctext{codec\_concatenated} & Meta-codec that concatenates a sequence of codecs (with intermediate mappers) \\
   \hline
   \ctext{blockmodem}%
      \footnote{The encoders/decoders in this section are implemented using
      the \ctext{blockmodem} interface due to the required access to the
      \ctext{channel}, which is only available through the \ctext{blockmodem}
      interface.}
   & \ctext{dminner} & Sparse inner codes with distributed marker sequence and Davey-MacKay decoder \cite{dm01ids} \\
   & \ctext{marker} & Marker codes with bit-level MAP decoder \cite{ratzer05telecom} \\
   & \ctext{tvb} & Time-Varying Block codes with GPU-enabled symbol-level MAP decoder \cite{bsw10icc,briffa13jcomml} \\
   \hline
   \ctext{mapper}
   & \ctext{map\_straight} & Each modulation symbol encodes exactly one encoder symbol \\
   & \ctext{map\_interleaved} & Random interleaving of symbols within the block \\
   & \ctext{map\_permuted} & Random permutation of symbols at each index \\
   & \ctext{map\_aggregating} & Each modulation symbol encodes more than one encoder symbol (e.g.\ binary codecs on $q$-ary modems) \\
   & \ctext{map\_dividing} & Each encoder symbol is represented by more than one modulation symbol (e.g.\ $q$-ary codecs on binary modems) \\
   & \ctext{map\_stipple} & Punctured mapper for turbo codes, with all information symbols transmitted and parity symbols taken from successive sets; equivalent to odd/even puncturing for two-set turbo codes \\
   & \ctext{map\_concatenated} & Meta-mapper that concatenates a sequence of mappers \\
   \hline
   \ctext{blockmodem}
   & \ctext{direct\_blockmodem} & Abstract $q$-ary channel modulation \\
   & \ctext{mpsk} & $M$-ary Phase Shift Keying modulation with Gray code mapping of adjacent symbols in constellation \\
   & \ctext{qam} & Quadrature Amplitude Modulation for square constellations with Gray code mapping of adjacent symbols \\
   \hline
   \ctext{channel}
   & \ctext{awgn} & Additive White Gaussian Noise channel (for signal-space modulations) \\
   & \ctext{laplacian} & Additive Laplacian Noise channel (for signal-space modulations) \\
   & \ctext{qec} & $q$-ary erasure channel (for abstract modulations) \\
   & \ctext{qsc} & $q$-ary symmetric substitution channel (for abstract modulations) \\
   & \ctext{qids} & $q$-ary insertion, deletion, and substitution channel (for abstract modulations) \\
   & \ctext{bpmr} & Bit-Patterned Media Recording channel of \cite{iyen11} \\
   }
This is followed by a list of codec sub-components in Table~\ref{tab:turbo}.
\inserttabd{tab:turbo}{ccp{5in}}
   {Summary of available codec sub-components.}{
   \emph{Base} & \emph{Class} & \emph{Description} \\
   \hline
   \ctext{fsm}%
      \footnote{This class implements a finite state machine interface,
      which is used to specify the encoder in a convolutional code.  This is
      needed for obvious reasons in the \ctext{mapcc} and \ctext{turbo}
      classes; it is also used to specify the accumulator in \ctext{repacc}
      and the mapping in \ctext{memoryless}.}
   & \ctext{dvbcrsc} & Circular recursive systematic convolutional code from the DVB standard \cite{dvb-rcs} \\
   & \ctext{gnrcc} & Non-recursive convolutional code over $\gf{q}$ with encoder polynomials expressed in controller-canonical form \\
   & \ctext{grscc} & Recursive convolutional code over $\gf{q}$ with encoder polynomials expressed in controller-canonical form \\
   & \ctext{nrcc} & Binary non-recursive convolutional code with encoder polynomials expressed in controller-canonical form \\
   & \ctext{rscc} & Binary recursive convolutional code with encoder polynomials expressed in controller-canonical form \\
   & \ctext{zsm} & A zero-state machine, or in other words a repeater \\
   & \ctext{cached\_fsm} & Meta-fsm that pre-computes and caches the input/output table of its component \ctext{fsm} \\
   \hline
   \ctext{interleaver}%
      \footnote{This class implemented the interface for the interleaver
      used in parallel concatenated convolutional codes, creating diversity
      between the parity sequences.}
   & \ctext{flat} & Null interleaver (usually used for the first parity sequence) \\
   & \ctext{berrou} & The original turbo code interleaver of \cite{berr93} \\
   & \ctext{helical} & Helical interleaver \cite{barb95} \\
   & \ctext{rectangular} & Simple rectangular interleaver \\
   & \ctext{shift\_lut} & Circular-shifting interleaver \\
   & \ctext{named\_lut} & A general interleaver specified as a look-up table (generated externally) \\
   & \ctext{uniform\_lut} & Random interleaver (with uniform distribution) \\
   & \ctext{rand\_lut} & Random interleaver with simile property \cite{barb95} \\
   & \ctext{onetimepad} & Interleaver that performs symbol-by-symbol modular addition between input and a random sequence \\
   & \ctext{padded} & Meta-interleaver that concatenates any interleaver with a \ctext{onetimepad} \\
   }

\subsection{The Monte Carlo simulator model}

In addition to the ease of setting up different combinations of codecs,
mappers, modems and channels, the framework also provides a feature-rich
simulator.
This can be configured to gather numerous performance measurements of the
described system, including SER, FER, and the timings of various components.
An overview of the simulator model implemented by our framework is
shown in Fig.~\ref{fig:sim_model}, where the communication system is
considered as a block box that performs the complete cycle of events of
Fig.~\ref{fig:comm_model}.
\insertfig{fig:sim_model}{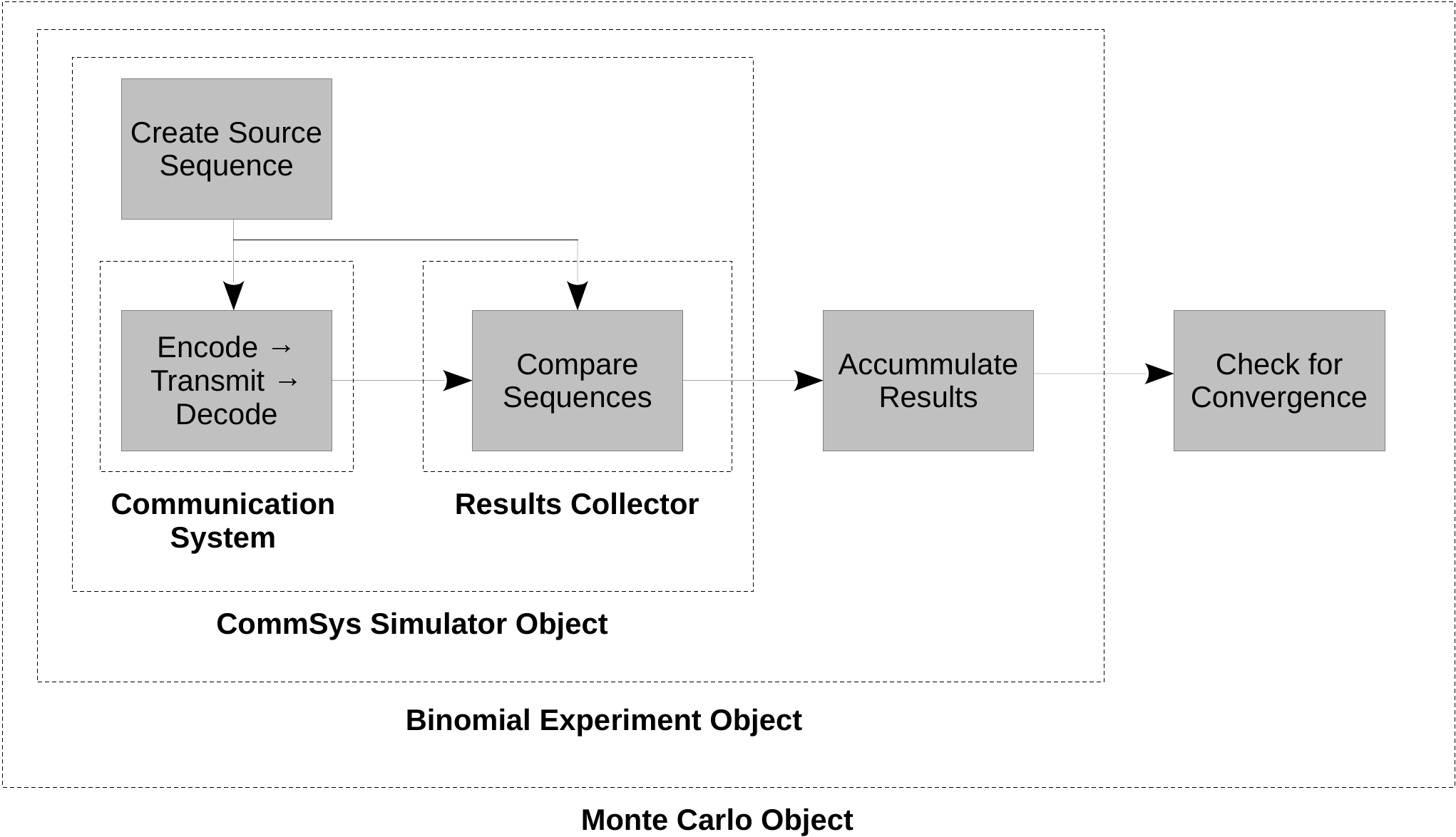}
   {The SimCommSys simulator model}
In SimCommSys, the simulator object defines the input sequences to be cycled
through the communication system object; a results collector component
compares the input and output of the communication system and computes the
required statistics.
This modular architecture allows the user to simulate any given communication
system under different input conditions and to collect a range of possible
results.
The simulator object implements a substantial part of the experiment interface:
that concerned with computing a single sample.
The accumulation of aggregate statistics from multiple samples is typically
performed by the binomial experiment object; this is suitable for error-rate
experiments, which can be seen as Bernoulli trials.
Finally, the Monte Carlo object implements the necessary loops to obtain
enough samples until convergence is achieved.
A summary of classes providing simulation types and related facilities
(such as results collectors) can be found in Table~\ref{tab:simulators}.
\inserttabd{tab:simulators}{ccp{3.75in}}
   {Summary of classes providing simulation types and related facilities
   (such as results collectors).}{
   \emph{Base} & \emph{Class} & \emph{Description} \\
   \hline
   \ctext{experiment\_binomial}%
      \footnote{Implements the interface for an experiment where a binomial
      proportion is estimated, approximating the error with a normal
      distribution.}
   & \ctext{commsys\_simulator} & Generic simulator of communication systems, supporting random, all-zero, or user-specified input sequences, and a modular results collection interface; simulation parameter specifies channel conditions \\
   & \ctext{commsys\_stream\_simulator} & Variation on \ctext{commsys\_simulator} that simulates stream transmission and reception, where the start and end of each frame are not assumed to be known \emph{a priori} and are instead estimated by the receiver \\
   & \ctext{commsys\_threshold} & Variation on \ctext{commsys\_simulator} where the simulation parameter specifies the modem threshold setting; the channel conditions are fixed to a value specified in this object \\
   \hline
   \ctext{experiment\_normal}%
      \footnote{Implements the interface for an experiment where the samples
      take a normal distribution, and the mean of the distribution needs to
      be estimated.}
   & \ctext{commsys\_timer} & Meta-experiment to determine timings of individual components of a given communication system \\
   \hline
   results collector%
      \footnote{While there is no base class, the interface is specified in
      \ctext{commsys\_simulator}, where objects of this type are used to
      specify the results to be collected in a given simulation.}
   & \ctext{errors\_hamming} & Conventional symbol and frame error rates (SER, FER) computed using the Hamming distance metric \\
   & \ctext{errors\_levenshtein} & As for \ctext{errors\_hamming}, with additional symbol error rate computed using the Levenshtein metric \cite{leve66} \\
   & \ctext{fidelity\_pos} & Computation of the fidelity metric at frame and codeword boundary positions, for synchronization error correcting codes \\
   & \ctext{hist\_symerr} & Computes histogram of symbol error count for each block simulated \\
   & \ctext{prof\_burst} & Determines the error probabilities for the first symbol in a frame, a symbol following a correctly-decoded one, and a symbol following an incorrectly-decoded one; used to determine the error burstiness profile \\
   & \ctext{prof\_pos} & Computes symbol-error histogram as dependent on position within block \\
   & \ctext{prof\_sym} & Computes symbol-error histogram as dependent on source symbol value \\
   }

\subsection{Local or distributed simulation}

The simulator (through the Monte Carlo object) is designed to work in a
distributed setup using a client-server model, but can also work as a single
local process.
The latter mode is useful, for example, for quick simulations or when
gathering timings for benchmarking and optimization.


In client-server mode, a server process controls the simulation and is
responsible for gathering all the results.
When starting a distributed simulation, the user specifies the port the
server listens on.
The user also passes a text-based configuration file that specifies the details
of the components needed in the simulation (see Section~\ref{sec:using}
for examples).
The user can then start any number of client processes either on the same
machine or across any number of networked computers.
The clients communicate with the server process using TCP/IP socket connections
to the specified port.
The server process keeps track of the number of clients and can accept new
ones at any time.
Additionally, if a client process dies or is otherwise disconnected, the
server removes it from the list of connections and is otherwise unaffected.
Together, these allow the user to dynamically scale the resources allocated
to a particular simulation.
Once the server process has finished the simulation, it will cleanly terminate
all the clients before stopping.
Should the server process stop inadvertently, for example if the user
switched off the machine running the server process, all clients terminate
automatically when their connection with the server process is lost.
This behaviour prevents the client processes from using up resources on the
client machines unneccessarily.
Fig.~\ref{fig:client_server_model} illustrates this principle.
\insertfig{fig:client_server_model}{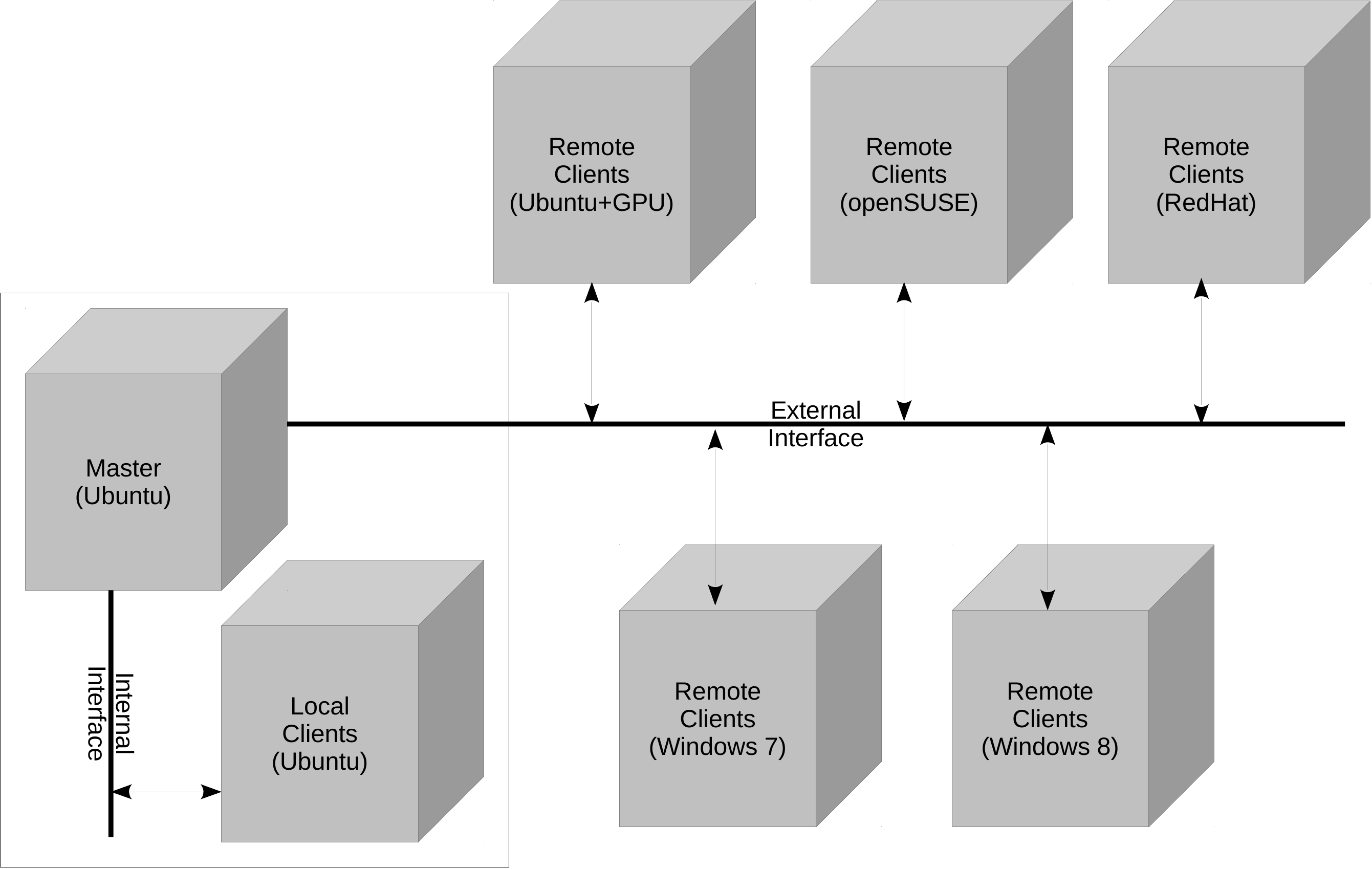}
   {The SimCommSys client-server model}
Note that in the distributed case, timings are only useful provided all the
client machines are homogeneous both in hardware and software.


When starting a simulation, the user also specifies the range of channel
conditions to be simulated and the convergence requirements for the simulation.
The channel conditions are usually specified in terms of the SNR or error
probabilities, with the range specified by the initial and final values and
a step factor.
Depending on the requirements of the channel used, the user can specify whether
the step factor is applied additively (e.g.\ in the case of the AWGN channel)
or multiplicatively (e.g.\ for abstract channels like insertion-deletion or
erasure channels).
Convergence requirements may be specified as the number of error events to
be accumulated (conventionally 100) or by specifying the required confidence
interval (as an error margin together with a confidence value).
When a confidence interval is set, the simulation is considered to have
converged when the true value is within the error margin of the current
estimate, at the stated confidence level.


Once the server process is up, all that any client process requires is
the hostname or IP address of the server process and the port number it is
listing on.
On connection, the server sends the system configuration to the client together
with the channel parameter to simulate.
Each client seeds its random generator with a true random value from the OS.
This ensures that all clients simulate different random input sequences, noise
patterns, and (for applicable systems) different random system components
(e.g.\ random interleavers).
The client instantiates all the necessary components and runs the simulation
sending results back to the server process regularly.
The server process aggregates results from all clients, and stores these
in a human-readable text file.
Intermediate results (i.e.\ aggregate results that have not yet converged)
are stored regularly on file, with the full state of the simulation.
This allows the server to continue a previously terminated simulation, and
mitigates the risk of server failure on long-running simulations.
The code base provides a python module and a simple script which can be
easily modified to present the results graphically using the matplotlib
library \cite{hunter2007}.

In all of the above the only programming required by the user is the
implementation of any new components and a simple adaptation of the python
script to visualise the results if required.
All the other aspects of the simulation are taken care of by the framework.
In the next section, we will demonstrate how to set up a system and run
a simulation.


\section{Using SimCommSys}
\label{sec:using}

\subsection{Running a quick simulation with a simple codec}

The most common use case is to set up and simulate a communication system
under a range of channel conditions.
This starts with the specification of a simulation based on a communication
system, defining the mapper, modem, and channel to be used as well as what
results we want to gather.
The file shown in Fig.~\ref{lst:uncoded-system} is an example of how to
achieve this for a very simple setup: an uncoded BPSK transmission over AWGN.
\insertsys{lst:uncoded-system}{Listings/uncoded-system.txt}
   {Uncoded BPSK transmission over AWGN}
As can be seen from the config file, each component has its own heading
followed by a number of parameters.
A version number is usually included to allow old configuration files to
be read provided the newer version of the serialisation code can provide
default values for any missing or changed parameters.
The file format allows the inclusion of comments, indicated by lines starting
with a \cverb|#|, which are skipped when the file is read.

In this example, we are simulating a communication system with a signal-space
channel representation and gathering error rates using the Hamming distance.
Within the simulator we can specify whether an all-zero, random or
user-specified information sequence should be used.
The actual communication system starts with the \cverb|uncoded| class defining
a binary code of length 16\,320 bits.
This is passed through a straight mapper (i.e.\ left unmodified) and modulated
with a PSK modem of alphabet size 2 (i.e.\ a BPSK modem).
The result is transmitted on an AWGN channel.
Note that the communication system allows us to define separate channel objects
for the transmit and receive functions.
This is unused here, and would allow us to simulate the system with a
mismatched receiver.

The user can test the system file by simply running a short simulation using
the `QuickSimulation' command as shown in Fig.~\ref{cmd:quicksimulation}.
\begin{lstcommand}{cmd:quicksimulation}{Running a short simulation}
quicksimulation.master.release -t 10 -r 6.8 -i Simulators/errors_hamming-random-awgn-bpsk-uncoded.txt >Results/sim.errors_hamming-random-awgn-bpsk-uncoded.txt
\end{lstcommand}
This will run the simulation at an SNR of $6.8\dB$ for ten seconds.
The final output should look something like Fig.~\ref{lst:uncoded-quicksim}.
\insertsys{lst:uncoded-quicksim}{Listings/uncoded-quicksim.txt}
   {Simulation output for uncoded BPSK transmission over AWGN}
This output also allows us to determine the speed at which a simulation
of this code runs.
For this simple codec, the simulator computed 135.8 frames of 16\,320 bits
each per second, equivalent to 2.22\,Mbit/s (on an Intel Core i5-3570K CPU
using a single core at 3.4\,GHz).

\subsection{A quick simulation of a more complex system}
\label{sec:concatenated}

Consider next a more complex system with a Reed-Solomon code in concatenation
with a convolutional code, as used in the NASA Voyager mission \cite{ms1993,
miller1981}.
The corresponding configuration file is shown in
Fig.~\ref{lst:concatenated-system}.
\insertsys{lst:concatenated-system}{Listings/concatenated-system.txt}
   {Serially concatenated Reed-Solomon and convolutional codes}
This defines a serially concatenated construction, as follows.
The outer code is a $(255,223)$ Reed-Solomon code defined over
$\mathbb{F}_{256}$ (using the \cverb|reedsolomon| component).
A random interleaver operates over four successive outer codewords, so that
any burst errors are distributed across four codewords (achieved using the
\cverb|map_interleaved| component).
The inner code is a rate-$\frac{1}{2}$ convolutional code, specified by the
feedforward polynomials $1 + z^{-2} + z^{-3} + z^{-5} + z^{-6}$ and
$1 + z^{-1} + z^{-2} + z^{-3} + z^{-6}$.
In the serialized file, these are represented by the strings 1011011 and
1111001.
The Reed-Solomon encoder output is converted from $\mathbb{F}_{256}$ to
$\mathbb{F}_{2}$ (binary) using the \cverb|map_dividing| component.
The output of four Reed-Solomon codewords is equivalent to 8\,160 bits; this is
terminated with six tail bits before encoding with the convolutional code.
The \cverb|mapcc| component uses the BCJR algorithm \cite{bahl74}, minimizing
SER.
The output for this code at an SNR of $2.2\dB$ after ten seconds is shown in
Fig.~\ref{lst:concatenated-quicksim}.
\insertsys{lst:concatenated-quicksim}{Listings/concatenated-quicksim.txt}
   {Simulation output for serially concatenated Reed-Solomon and convolutional codes}
Note that this corresponds to a decoding speed of 36.9\,kbit/s which is about
sixty times slower than the uncoded transmission.

\subsection{Running a proper simulation}

Having tested the simulator configuration files and verified that they are
working, it is now a simple step to start a simulation.
For the uncoded system, this can be done as shown in Fig.~\ref{cmd:server}.
\begin{lstcommand}{cmd:server}{Running the server process for a proper simulation}
simcommsys.master.release -i Simulators/errors_hamming-random-awgn-bpsk-uncoded.txt -o Results/errors_hamming-random-awgn-bpsk-uncoded.txt -e :9000  --start 1.5 --stop 11 --step 0.1 --floor-min 1e-5 --confidence 0.95 --relative-error 0.05
\end{lstcommand}
This starts the server process, listening on port $9000$.
The simulation starts with the channel SNR at $1.5\dB$ and works its way up
to $11\dB$ in steps of $0.1\dB$.
The point at which the simulation switches to the next SNR value is determined
by the convergence requirements, which can be specified either in terms of
the number of error events encountered or as a confidence interval.
In this case we specify an error margin of $\pm5\%$ at a $95\%$ confidence
level.

In this example, the simulation stops completely when at least one of the
the measures (SER or FER in this case) has fallen below $10^{-5}$.
Note that this can and usually should occur well before all noise values have
been explored.
Alternatively, the user can set \cverb|--floor-max| which would require all
measures to fall below this threshold before the simulation stops.
As SER $\leq$ FER, `floor-min' is usually used if the SER is more important to
measure while `floor-max' is used if the FER is the main focus.

Any number of clients can now be started using the command shown in
Fig.\ref{cmd:client},
\begin{lstcommand}{cmd:client}{Running the client process}
simcommsys.master.release -e server_address:9000
\end{lstcommand}
where the \cverb|server_address| could be \cverb|localhost| if the client
is started on the same machine, or alternatively the IP address or the
DNS-resolvable name of the computer where the server process is running.

Repeating the same process for the concatenated system of
Section~\ref{sec:concatenated} and for a system with the inner convolutional
code alone, we now have the necessary results to plot a performance comparison,
as shown in Fig.~\ref{fig:nasa-ber}.
\insertfig{fig:nasa-ber}{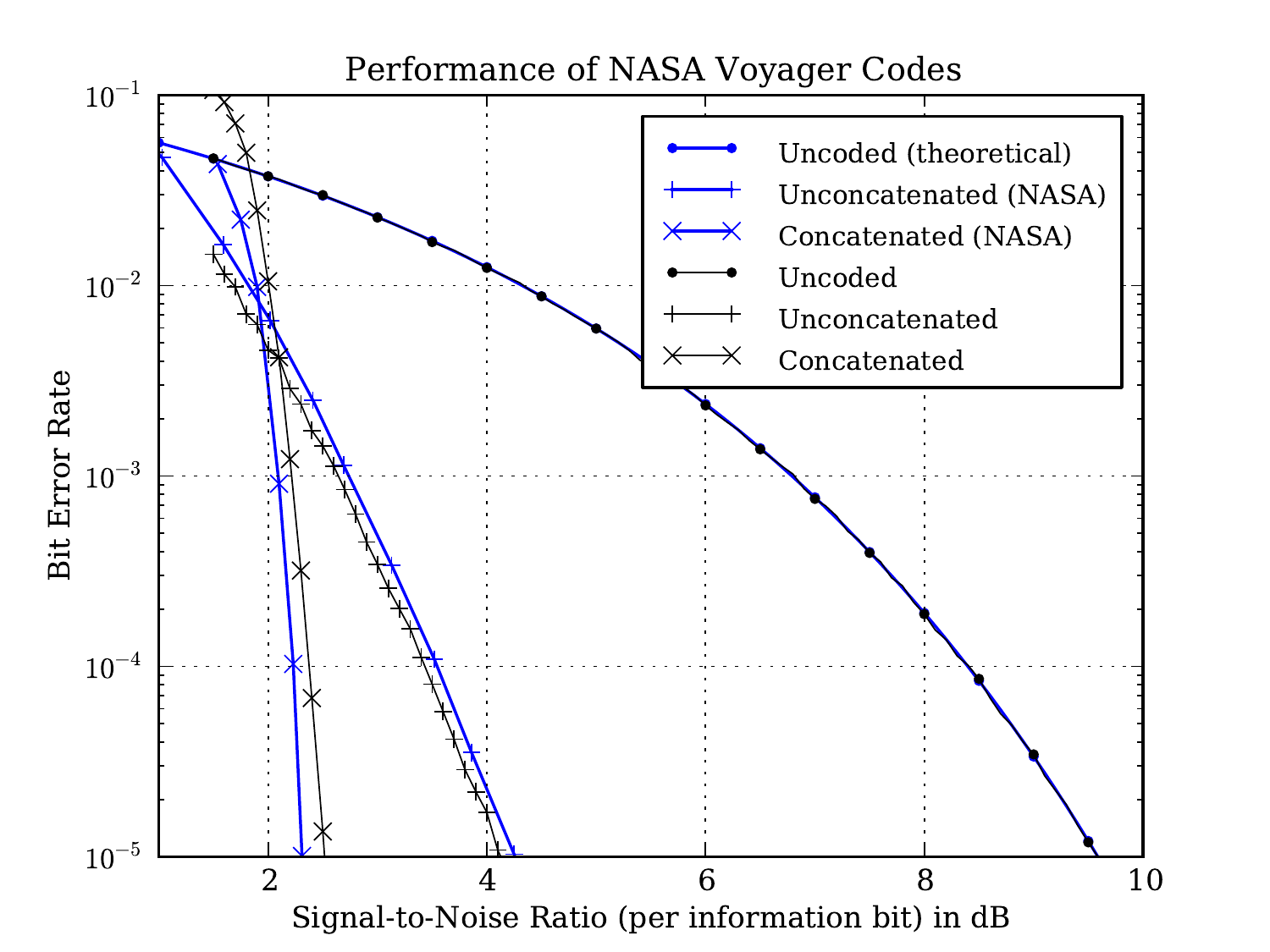}
   {Performance comparison of two codes used in the NASA Voyager mission.
   Previously published results from \cite{miller1981} and theoretical results
   for the uncoded channel are shown in blue.}
The theoretical error rate for the uncoded AWGN channel is also shown, clearly
coinciding with our simulation.
For comparison, the figure also includes previously published results from
\cite{miller1981}.
As expected, our simulation of the same convolutional code shows better
performance; this is because we use a bit-optimal (MAP) decoder rather than
the Viterbi decoder.
On the other hand, our simulation of the concatenated system shows slightly
worse performance since we use a random interleaver covering four Reed-Solomon
codewords rather than an infinite interleaver.

A collection of example systems is included in the repository.
This includes the systems shown in this section, together with the simulation
results and plot script.
For further details on constructing simulations for specific components,
please consult the user documentation.


\section{Extending SimCommSys}
\label{sec:extending}

To illustrate how the framework can be extended we will implement the simplest
of codes, the `uncoded' code, where the information sequence is transmitted
as is and the received sequence is simply decoded by taking a hard decision.
This will illustrate several features and concepts of the framework without
getting bogged down in any implementation issues of an actual codec.
To make this exercise slightly more interesting, our implementation extends
the \cverb|codec_softout| abstract class rather than the plain \cverb|codec|
abstract class.
This class defines the interface for a soft-input, soft-output codec, and
allows us to look at some of the features used by most modern codes.

The main method that needs to be implemented for the encoding is the
\cverb|do_encode| method of Fig.~\ref{lst:uncoded-encode}.
\begin{lstcode}{lst:uncoded-encode}{uncoded.cpp (encode)}
template <class dbl>
void uncoded<dbl>::do_encode(const array1i_t& src, array1i_t& enc)
   {
   // Copy input to output
   enc = src;
   }
\end{lstcode}
Note the use of templates in the code fragment.
Throughout the framework, templates are used to provide support for a range of
code alphabets and numerical precisions with very little extra programming
effort.
The actual code alphabet and numerical precision are defined when the
templates are instantiated, and can be chosen at run-time as part of the
serialization process.

Decoding is implemented as a two-step process.
First, the decoder in initialized with the probabilities of the received
sequence.
For a soft-input, soft-output codec, the methods that need to be
implemented for this are the \cverb|do_init_decoder| methods of
Fig.~\ref{lst:uncoded-initdecoder}.
\begin{lstcode}{lst:uncoded-initdecoder}{uncoded.cpp (init\_decoder)}
template <class dbl>
void uncoded<dbl>::do_init_decoder(const array1vd_t& ptable)
   {
   // R is a private variable of the uncoded class
   R = ptable;
   }

template <class dbl>
void uncoded<dbl>::do_init_decoder(const array1vd_t& ptable, const array1vd_t& app)
   {
   // Initialize results to received statistics
   do_init_decoder(ptable);
   // Multiply with prior statistics
   R *= app;
   }
\end{lstcode}
Note that this method is overloaded, having two implementations with different
parameters.
Therefore, two variants of this method need to be implemented:
one where only the channel statistics for the received sequence are available,
and another where prior probabilities for the transmitted sequence are
also given.
The latter interface is required for systems involving iteration between
the \cverb|modem| and \cverb|codec| components.

Next, the actual decoding takes place; this may be repeated a number of times
in an iteratively decoded code, where each successive decoding makes use of
information from previous decodings.
For a soft-input, soft-output codec, the methods that need to
be implemented for this are the \cverb|softdecode| methods of
Fig.~\ref{lst:uncoded-softdecode}.
\begin{lstcode}{lst:uncoded-softdecode}{uncoded.cpp (decode)}
template <class dbl>
void uncoded<dbl>::softdecode(array1vd_t& ri)
   {
   // Set input-referred stats to stored values
   ri = R;
   }

template <class dbl>
void uncoded<dbl>::softdecode(array1vd_t& ri, array1vd_t& ro)
   {
   // Set input-referred stats to stored values
   ri = R;
   // Set output-referred stats to stored values
   ro = R;
   }
\end{lstcode}
This method is also overloaded:
the first implementation computes the posterior probabilities of the decoded
sequence only, while the second one also computes the posterior probabilities
of the encoded sequence.
It is up to the calling class to decide which of these methods to use as there
may be a computational cost in computing both values when only one is needed.
For example, in the case of turbo codes, only the probabilities of the
information symbols are required as input to the next decoding stage while
the probabilities of the parity check symbols are not required.
On the other hand, with LDPC codes the Sum-Product Algorithm needs to compute
the probability of all symbols as everything is used in the next iteration.

The last two required functions provide serialisation support to the codec.
These allow the codec to be read in from file when setting up a simulation,
and also allow the server to send a serialised version of the system to be
simulated to its clients.
To achieve this, each component of the system needs to implement the
following two methods of the \cverb|serializable| abstract class.
In this example, the \cverb|serializable| interface is inherited through the
\cverb|codec| and thus the \cverb|codec_softout| class.
The first method writes the details of the codec to an output stream
(e.g.\ a file or a socket), while the second method reads the necessary
class parameters (and components, where applicable) from an input stream.
Both are shown in Fig.~\ref{lst:uncoded-serialize}.
\begin{lstcode}{lst:uncoded-serialize}{uncoded.cpp (serialize)}
template <class dbl>
std::ostream& uncoded<dbl>::serialize(std::ostream& sout) const
   {
   sout << "# Version" << std::endl;
   sout << 1 << std::endl;
   sout << "# Alphabet size" << std::endl;
   sout << q << std::endl;
   sout << "# Block length" << std::endl;
   sout << N << std::endl;
   return sout;
   }

template <class dbl>
std::istream& uncoded<dbl>::serialize(std::istream& sin)
   {
   // get format version
   int version;
   sin >> libbase::eatcomments >> version;
   // read the alphabet size and block length
   sin >> libbase::eatcomments >> q >> libbase::verify;
   sin >> libbase::eatcomments >> N >> libbase::verify;
   return sin;
   }
\end{lstcode}
Note that the \cverb|libbase:eatcomments| manipulator filters out any comments
in the input stream.
There are some more boiler-plate methods that need to be implemented,
including methods to return a short description of the codec and methods to
return the length, dimension, and alphabet size of the code.

For such a templated class, the implementation file needs to contain explicit
instantiations of the different combinations of template parameters of
the codec.
Fig.~\ref{lst:uncoded-boost} shows how Boost preprocessor metaprogramming
\cite{ag2004} is used to create the various instances of the templated codec
which can then be serialized using the name of the class.
\begin{lstcode}{lst:uncoded-boost}{uncoded.cpp (boost)}
namespace libcomm {

// Explicit Realizations
#include <boost/preprocessor/seq/for_each.hpp>
#include <boost/preprocessor/stringize.hpp>

using libbase::serializer;
using libbase::mpreal;
using libbase::mpgnu;
using libbase::logreal;
using libbase::logrealfast;

#define REAL_TYPE_SEQ \
   (float)(double) \
   (mpreal)(mpgnu) \
   (logreal)(logrealfast)

/* Serialization string: uncoded<real>
 * where:
 *      real = float | double | mpreal | mpgnu | logreal | logrealfast
 */
#define INSTANTIATE(r, x, type) \
   template class uncoded<type>; \
   template <> \
   const serializer uncoded<type>::shelper( \
      "codec", \
      "uncoded<" BOOST_PP_STRINGIZE(type) ">", \
      uncoded<type>::create); \

BOOST_PP_SEQ_FOR_EACH(INSTANTIATE, x, REAL_TYPE_SEQ)

} // end namespace
\end{lstcode}
Finally, we must ensure that serialization system `sees' an object of this
type on startup: this is done by adding an object of this type as a field
or parent class of \cverb|serializer_libcomm|.
For templated classes, it is enough to include an object with any of the
explicitly instantiated template parameters.


\section{Conclusion}
\label{sec:closure}

In this paper we have presented SimCommSys, a Simulator of Communication
Systems released under an open-source license.
An overview was given of the core of the project, a set of C++ libraries
defining a number of communication system components and a distributed Monte
Carlo simulator.
The more common use case, where a communication system is defined and then
simulated using the framework, was demonstrated.
Finally, a tutorial for extending the framework was given, using the
implementation of an `uncoded' codec as an example.

SimCommSys fills a current void, providing a reliable platform for
simulating communication systems without any additional programming
(as long as the required components are already available in SimCommSys).
Development is only necessary when extending the framework by implementing
new components.
The strict separation of development and the framework's use to simulate
specific constructions encourages reproducibility of experimental work and
reduces the likelihood of error.

The project has been in development for many years, and has evolved to
support changing requirements.
For example, most recently the framework has been used to simulate codes for
synchronization error channels, which as far as we know is not supported by
any other publicly available software.
We expect the project to continue to evolve as the needs of its users change.
The wider its user base, the more comprehensive the software will become.
We encourage potential users to download and use the project in their own
research.
Support is available through the project forum, while feature requests and
bug reports may be submitted through the project tracker.
Developers who wish to contribute to the project are asked to contact the
maintainer; community support is welcome.


\balance

\bibliographystyle{IEEEtran}
\bibliography{IEEEabrv,jbabrv,jb,turbo,unsorted,victor}

\end{document}